\documentstyle[aps,prl,twocolumn,floats,epsfig,amssymb]{revtex}

\begin{document}
\draft \wideabs{
\title{Tunable Resonant Optical MicroCavities by Self-Assembled Templating}
\author{ G.V. Prakash,$^{1}$  L. Besombes,$^{1}$ T. Kelf,$^{1}$ P.N. Bartlett,$^{2}$
M.E. Abdelsalam,$^{2}$ and J.J. Baumberg$^{1}$}
\address{$^{1}$School of Physics \& Astronomy, University of Southampton, Southampton, SO17 1BJ, UK}
\address{$^{1}$School of Chemistry, University of Southampton, Southampton, SO17 1BJ, UK}
\date{\today}
\maketitle
\begin{abstract}
Micron-scale optical cavities are produced using a combination of template sphere
self-assembly and electrochemical growth. Transmission measurements of the tunable
microcavities show sharp resonant modes with a $Q$-factor$>$300, and 25-fold local
enhancement of light intensity. The presence of transverse optical modes confirms
the lateral confinement of photons. Calculations show sub-micron mode volumes are
feasible. The small mode volume of these microcavities promises to lead to a wide
range of applications in microlasers, atom optics, quantum information, biophotonics
and single molecule detection.
\end{abstract}
\vspace{-0.2cm}  }

\vspace{-1cm} \narrowtext

The operation of every type of laser, as well as many varieties of atomic traps and
optical sensors, depend on the formation of an optical cavity. The resonant optical
modes inside this cavity determine the spatial field distribution and spectral
performance of the device. The most common optical designs use spherical mirrors to
form a confocal cavity that is insensitive to misalignment and confines the optical
mode to the smallest lateral dimensions.\cite{Seigman} The length of this cavity is
on the order of the radius of curvature of the mirrors. There has been much interest
in the problem of reducing the dimensions of such cavities: planar microcavities
have been widely used as a way to control spontaneous emission and to enhance the
interaction of light with matter, as in quantum wells\cite{Khitrova} or quantum dots
\cite{Saito} but these structures only confine photons in one dimension. Confinement
in lateral dimensions such as in photonic crystals\cite{Yablonovitch} or microcavity
mesas\cite{Gerard96,Gerard99} can inhibit spontaneous emission altogether, but
involve complex and expensive fabrication strategies. Here we present a simpler
approach utilising confocal microcavities. While traditional lasers built from
discrete components use macroscopic spherical mirrors, so far microcavity lasers do
not. Preliminary theoretical work on the mode structure in parabolic
dome\cite{Nockel} and spherical cavities\cite{Abram} highlight the promise of such
0D microcavities, but confirm that fabrication is nontrivial. Similarly glass or
polymer microspheres show high Q-factors in whispering gallery modes but it is
generally hard to control them and to couple light into and out of them.\cite{Jia}

To construct cavities with the smallest possible mode volume requires small
radius-of-curvature mirrors. Using our recently devised route to simple fabrication
of such spherical micro-mirrors, we demonstrate here the formation of stable
microcavities on size-scales below 10 microns. Comparable vertical-cavity
surface-emitting semiconductor lasers use planar mirror stacks integrated into the
layered semiconductor growth, and then require sensitive lateral fabrication to
produce small volume devices.\cite{Gerard96,Jewell,Lear} Comparable whispering
gallery modes in micro-spheres or micro-disks have larger mode volumes and broadband
tuning of the resonant cavity wavelengths is a severe problem. By robustly embedding
our spherical micro-mirrors in thin metal films, we obtain simple voltage tuning by
piezo-electric translation of a top planar mirror. We thus show that at
length-scales approaching the wavelength of light, the electric field distribution
in such microcavities can be controlled by nanoscale tuning of the mirror, and can
still be interpreted using Gaussian beam optics.
\begin{figure}[t]
\centering\epsfig{file=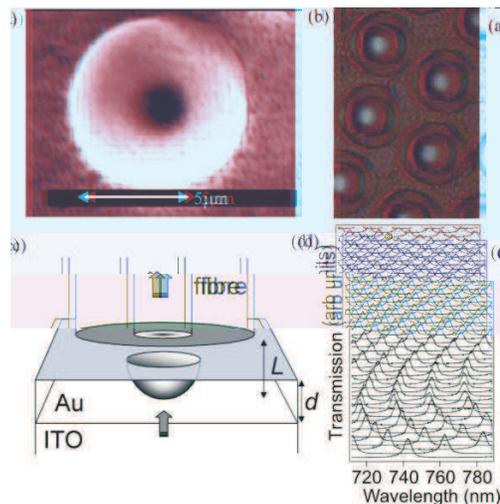,width=6.5cm}
\caption{(a) SEM and (b) optical micrograph of spherical micro-mirror of pore
diameter 8$\mu$m, radius of curvature $R$=5$\mu$m, in 2.5$\mu$m-thick Au on ITO. (c)
Microcavity design. (d) Microcavity transmission spectra collected through the fibre
for increasing separation, $L$, of cavity mirrors (spectra displaced vertically
upwards, for each increase in cavity length $\sim$200nm).}
 \vspace{-0.3cm}
 \label{fig1}
\end{figure}

Until recently, making curved mirrors in the size range required has been difficult,
but advances in electrochemistry and directed self-assembly of colloidal templates
now offer a cheap and simple way to produce an array of well-ordered spherical
reflecting surfaces.\cite{Bartlett} For the purposes of constructing microcavities,
we combine spherical micro-mirrors of Au with radii $R<10\mu$m embedded in a thin
film, with a planar top Au mirror coated on a cleaved single-mode optical fibre. The
guided mode of the fibre is then also used to efficiently couple light out of the
cavity. Briefly, we use a template prepared through sedimentation of a confined
colloidal solution of latex spheres on an indium-tin-oxide (ITO) coated substrate
electrode to leave a self-assembled arrangement. Either close-packed arrays or
isolated sphere templates can be obtained by adjusting the sphere concentration or
by pre-patterning. The substrate is then placed in an electrochemical cell with a
solution of aqueous metal complex ions which are deposited through reduction in the
interstitial spaces of the template. The resulting film thickness can be controlled
precisely through measurement of the total charge, yielding cavity shapes that range
from shallow dishes to nearly-complete spherical voids. The latex spheres are
removed by dissolving in tetrahydrofuran, leaving a porous metallic `cast' with the
ordering and size of the original template. Both electron and optical microscopy
(Figs.1a,b) give an indication of the mirror quality, discussed
elsewhere.\cite{Coyle}

By growing the micro-mirrors on semi-transparent substrates, we take advantage of
the negligibly thin Au coating at the base of the mirror in order to couple light
into the cavity. The results shown here use radius-of-curvature mirrors, $R$=5 and
10$\mu$m grown up to a Au thickness of around 2$\mu$m to form dish-like spherical
reflectors between 5$\mu$m and 15$\mu$m across. However we have also successfully
grown spherical reflectors down to $R$=100nm radius-of-curvature. The upper cavity
mirrors are formed by evaporating a 28nm thick Au film (80$\%$ reflectivity at
$\lambda$=750nm) on a perpendicularly-cleaved stripped single-mode optical fibre
(SM660) of 100$\mu$m outer diameter, which is mounted in an {\it XYZ} piezoelectric
(PZT) translation stage and aligned normally to the micro-mirror film (Fig.1c). A
white-light source is focussed to a 10$\mu$m spot on the rear side of the
micro-mirror film, and co-linear imaging allows us to back-illuminate individual
spherical mirrors. As the fibre approaches the film, optimising the collected
transmission signal brings it into correct alignment with the micro-mirror, forming
the microcavity. The detected light emerging from the fibre is the product of the
cavity transmission and the coupling strength to the fundamental fibre mode. Using
the fibre as the top mirror thus allows a self-aligned detection geometry. We record
the spectra by directly coupling the output end of the fibre to a 0.5m monochromator
and CCD of combined 0.05nm resolution.
\begin{figure}[h]
\centering\epsfig{file=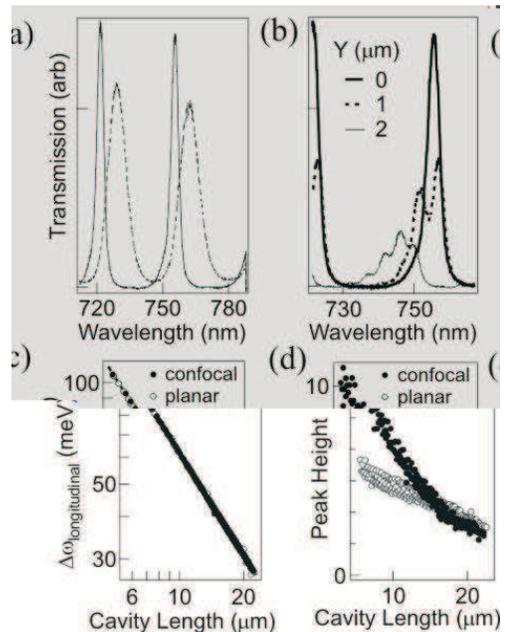,width=6.5cm}
 \vspace{0.05cm}
 \caption{Transmission
spectra for cavity lengths, $L\sim$7.9$\mu$m, $R$=10$\mu$m, for (a) plane (dashed)
and confocal (solid) microcavities, and for (d) confocal microcavity with the fibre
increasingly laterally shifted from the central axis ($Y$=0,1,2$\mu$m). (b)
Longitudinal cavity mode separation for planar and confocal cavities, together with
the prediction from Gaussian optics (line). (c) Peak cavity transmission as the
cavity length increases.
 }
 \vspace{-0.3cm}
 \label{fig2}
\end{figure}

Spectral characterisation of the optical transmission through the device allows us
to compare the microcavities with simple models based on Gaussian beam optics. We
also produce similar 28nm-thick plane Au reflectors to directly compare our results
to the plane-plane microcavity geometry using the same top fibre mirror. Typical
experimental results for the transmitted modes are seen in Fig.1d, and are stable
and reproducible, and similar in most well-formed spherical micro-reflectors. As
expected, the spacing of longitudinal modes becomes closer together as the cavity
length increases. The absolute transmission is only low here because the input
coupling through the sub-micron aperture is very small for the incident incoherent
white light source. Equivalent scans on a plane mirror reveal similar longitudinal
modes but with a much lower finesse. A direct comparison between the plane-plane
mirrors (a `planar' cavity) and plane-spherical mirrors (termed `confocal') is shown
in Fig.2a for a cavity length, $L\sim$7.9$\mu$m, with $R$=10$\mu$m. The general
formula for the resonant frequency, $\omega_{npl}$, that produces zero net optical
phase after one round trip is given by\cite{Seigman}
\begin{equation}
\omega_{npl} = \frac{c}{L} \{ \pi n - \theta_{Au} + (2p+l+1) \tan^{-1}
\sqrt{\frac{L}{R-L}} \}
\end{equation}
where $n,p,l$ are positive integers describing the longitudinal, transverse radial
and azimuthal mode numbers of the resonance, the second term represents the phase
shift on reflection from the metal film ($\theta_{Au}\simeq 180^{\circ}-24^{\circ}$
is the non-ideal phase shift on reflection for Au at these wavelengths), and the
third term is the Gouy phase shift arising from the range of propagation angles
contained in the Gaussian mode. From this paraxial approximation, the separation of
the longitudinal modes yields the cavity length directly. Plotting the experimental
and theoretical longitudinal mode splitting, $\Delta \omega_L$, as a function of the
cavity length derived from the calibration of the PZT $Z$ translation gives an
excellent fit for both types of cavity (Fig.2c).

The difference between the planar and confocal microcavities becomes clearer when
the transmission peak heights and linewidths are compared. At cavity lengths $L\! <
\! R$=10$\mu$m, the finesse of the confocal microcavity approaches 10, while the
quality-factor, $Q\! >\! 300$. On the other hand, the finesse of the planar
microcavity does not exceed 4 due to the diffraction losses of the optical modes.
This improvement can be directly seen from the peak transmission intensity vs.
cavity length (Fig.2d), which is strongly enhanced for mirror separations $< \! R$.

To prove that the spherical mirror is acting to localise the optical modes in the
transverse direction, the core of the optical fibre is laterally scanned in the {\it
XY}-plane across the top opening of the spherical micro-mirror at a fixed cavity
length. Typical spectra are shown in Fig.2b, and resolve a number of new optical
modes. These modes always appear on the short wavelength side of the original
longitudinal mode and thus can be identified as transverse modes formed by the
lateral optical confinement of the spherical micro-mirror. They are never seen in
plane-plane microcavities. To further study the spatial dependence of the spectral
transmission, we systematically collect spectra at different $X$ positions and plot
the resulting images. Scanning across the centre of the microcavity (Fig.3a, $Y$=0)
clearly shows the emergence of the transverse modes away from the cavity axis. The
higher transverse modes dominate the spectra for scans across a chord of the
circular opening (Fig.3b, $Y$=2$\mu$m). Up to 6 transverse modes with similar
linewidths make up each longitudinal mode. Scanning in the orthogonal direction
gives similar results, indicating a near-cylindrical symmetry of the
micro-resonator.
\begin{figure}[h]
\centering\epsfig{file=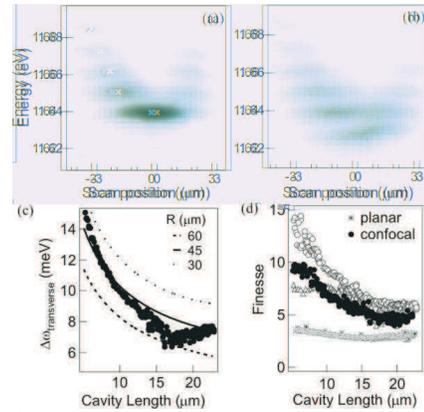,width=5.5cm} \vspace{0.02cm}
 \caption{Transmission spectra vs. lateral fibre $X$ position (log scale over 1.5 decades) for
(a) $Y$=0 and (b) $Y$=2$\mu$m, around the spectral position of a single longitudinal
mode for the confocal cavity (with $R$=10$\mu$m) at a cavity length of 7.6$\mu$m.
The white crosses in (a) are predicted positions of the transverse modes collected
by the fibre core. (c) Transverse mode separation vs. cavity length for $Y$=1$\mu$m.
(d) Cavity finesse vs. cavity length for planar ($\times$) and confocal
microcavities with one ($\bullet$) and two ($\circ,\bigtriangleup$) transverse
modes.
 }
 \vspace{-0.4cm}
 \label{fig3}
\end{figure}

These results thus allow one to directly image the different spatial mode
distributions inside the microcavity, and show that the mode spectrum consists of
discrete frequencies. An emitter inside a planar microcavity can always emit in some
particular direction - equation (1) is valid only for normal incidence, and at
different angles of incidence the resonance frequency gradually shifts to shorter
wavelengths. In fundamental contrast to this behaviour, an emitter in the confocal
microcavity does not have this flexibility and can only emit in the appropriate
field mode, $\omega_{npl}$. The transverse mode separation, $\Delta \omega_T$, also
decreases as the cavity length increases (Fig.3c), as expected from different values
of $(2p+l)$ in equation (1). However the predicted separations (Fig.3c, lines) only
account reasonably for the data if the radius-of-curvature, $R_{fit}$, of the
spherical micro-mirror is $\sim$45$\mu$m (as discussed below). Less pronounced
transverse modes are also seen in $R$=5$\mu$m microcavities, with similar transverse
splittings, $\Delta \omega_T$=12meV at $L$=8$\mu$m. The transverse spectral
separation produces a corresponding lateral mode spatial separation on the order of
a few hundred nanometres, providing a potentially effective way to couple into
specific cavity modes.

The finesse of the different spectral modes is shown in Fig.3d as a function of the
cavity length. Whereas the finesse of the planar structure is $\sim$4, independent
of the spacing between the mirrors, the confocal microcavity performs best at small
cavity separations. The maximum finesse of 15 is achieved for the lowest of the
transverse modes. This agrees well with the theoretical finesse of $\pi
\sqrt{r}/(1-r) \sim$14, where $r$=0.8 is the product of the amplitude reflectivity
coefficients of the two mirrors. The corresponding intensity concentrated within the
cavity is $(1-r)^{-2}\sim$25 times greater than that of the incident light at the
resonant wavelength. However, as the cavity length increases beyond the mirror
curvature $R$, the finesse drops. This is due to the increasing diameter of the
transverse field profile on the spherical micro-mirror, eventually leading to it
being clipped when $L>R$.

In this cylindrically symmetric micro-resonator, the field modes in the paraxial
approximation are given by Laguerre-Gauss functions,\cite{Seigman} plotted in
Fig.(4a) at the position of the top fibre mirror. This demonstrates how the
microcavity dimensions here are nearly optimal to couple cavity modes into the
lowest guided modes in the fibre. Higher order transverse modes have a larger
diameter on the cavity mirrors hence they suffer extra loss and exhibit lower
finesse, as seen in Fig.(3d). However asymmetric modes possess a field null on axis
that is useful for the optical trapping of dark field species.\cite{MacDonald}
\begin{figure}[h]
\centering\epsfig{file=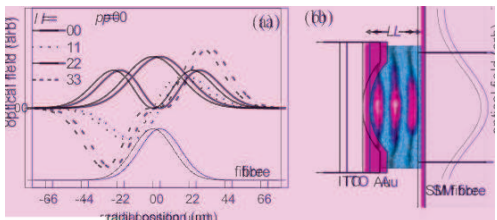,width=6.5cm}
 \vspace{0.01cm}
\caption{(a) Optical field profiles for the fibre and the first four Laguerre-Gauss
modes of the $R$=10$\mu$m confocal cavity with $p$=0. (b) Calculated field
distribution in a spherical confocal microcavity on resonance, $R$=880nm, $L$=
704nm.
 }
 \vspace{-0.3cm}
 \label{fig4}
\end{figure}

To understand the expected spatial patterns collected by the fibre core (with its
field distribution also illustrated in Fig.4a), we convolve the cavity and fibre
modes to calculate the mode overlap as a function of lateral fibre position. The
higher transverse modes are generally {\it not} seen for on-axial alignment due to
the averaging of the oscillating electric field by the larger extent of the fibre
mode. This is even more pronounced in the smaller ($R<$5$\mu$m) spherical
microcavities which give correspondingly smaller mode diameters. The calculated
lateral offsets which optimise collection of the different transverse modes are
plotted as white crosses in Fig.3a for $R$=45$\mu$m, giving excellent agreement with
the experimental data. This concurs with the need for a larger $R_{fit}$ that is
required to account for the transverse mode spectral separation, and suggests that
the spherical geometry of our micro-mirrors is imperfect. In fact, as is visible on
both the micrographs of Fig.1a,b and in scanning force microscopy images (not
shown), the central portion of the micro-mirrors is completely flat ($R=\infty$)
where the electrochemical growth appears to be screened under the template spheres.
The area of the flat region is about 10$\%$ of the total mirror area, a ratio which
remains approximately the same for spherical mirrors of radius down to 100nm. Thus
an effective cavity radius-of-curvature determined by both the curved and flat
sections is appropriate for the theory, in accord with the experimental results.
More detailed calculations based on realistic micro-mirror shapes cannot rely on the
formalism for Gaussian mode propagation. However, the close match between
experimental and theoretical finesse proves that this non-spherical mirror does not
degrade the phase front of the lowest cavity mode, and that the mirror flatness is
not a serious problem.

In 5$\mu$m radius-of-curvature cavities, the tight transverse field profiles (radius
$w$) imply diffraction angles $\theta \simeq \lambda / \pi w >$ 0.50 radians in the
cavity, leading to significant errors of order 5$\%$ in the paraxial approximation,
$\sin \theta = \theta$. As the size of such cavities is further reduced, these
errors rapidly make conventional optical models inappropriate. Theoretical modelling
to demonstrate the size scaling of these microcavities should therefore be carried
out using a full solution of Maxwells' equations in these wavelength-scale
geometries. The ultimate aim is to create the smallest optical cavities ($L \sim
\lambda/2$) with the highest finesse. A typical electric field distribution is shown
in Fig.4b for $R$=880nm, $L$=704nm, and $\lambda$=750nm with metallic reflectors (of
conductivity 200$/ \Omega m$ hence 80$\%$ reflectivity), corresponding to a mode
volume below (1$\mu$m)$^3$. This shows that our templating scheme is suitable for
sub-micron optical devices. For example, sub-picosecond optical modulators using
resonant cavity enhancement of nonlinearities require short cavity lifetimes and
hence short cavity lengths.

To summarise our key results, we have been able to fabricate near-spherical
micro-mirror optical cavities with mode volumes below (5$\mu$m)$^3$, a $Q \! > \!
300$ and finesse $>$10, and an intensity enhancement $>$25. Both longitudinal and
transverse optical modes are observed, and can be simply wavelength-tuned by PZT
translation of the planar top mirror which also extracts light from the cavity. In
addition, this top fibre mirror can be replaced by high-reflectivity multilayers
supporting luminescent nano-particles, such as semiconductor quantum dots, for
enhanced emission and microlasing. The new route this opens for cavity enhancement
of light-matter coupling can be further developed by reducing still further the
cavity volume (we routinely produce sub-micron reflectors) and improving the finesse
(using Ag or multilayer dielectric coatings), both now in progress. We note that
metal mirror cavities with similar finesse have already been shown to be sufficient
for attaining strong light-matter coupling in organic chromaphores.\cite{Hobson} Our
preliminary measurements already suggest that these micro-mirror cavities are useful
for optical-dipole force trapping, for enhanced collection of
micro-photoluminescence from the focal region, or for incorporation in experiments
involving cold atom interactions on integrated chips.

We gratefully thank Ed Hinds and Tim Freegarde for helpful discussions of resonator
design, and Oliver Wright and Dave Hanna for critical comments. We acknowledge
financial support from JSPS, EPSRC GR/N37261, GR/R54194, GR/S02662 and GR/N18598.

\end{document}